\documentclass[sigconf]{acmart}

\usepackage{booktabs}
\usepackage{subcaption}
\usepackage{listings}
\usepackage{threeparttable}
\usepackage{multirow}
\usepackage{mwe,tikz}
\usepackage{siunitx}
\usepackage{xcolor,colortbl}
\usepackage{xparse}
\usepackage{enumitem}

\newcommand{\name}{$f$\kern-0.18em\emph{unc}\kern-0.05em X}

\newcommand{\dlhub}{DLHub}
\newcommand{\skluma}{Skluma}

\newsavebox{\fminipagebox}
\NewDocumentEnvironment{fminipage}{m O{\fboxsep}}
 {\par\kern#2\noindent\begin{lrbox}{\fminipagebox}
  \begin{minipage}{#1}\ignorespaces}
 {\end{minipage}\end{lrbox}%
  \makebox[#1]{%
    \kern\dimexpr-\fboxsep-\fboxrule\relax
    \fbox{\usebox{\fminipagebox}}%
    \kern\dimexpr-\fboxsep-\fboxrule\relax
  }\par\kern#2
 }

 \definecolor{deepred}{rgb}{0.6,0,0}
 \definecolor{deepgreen}{rgb}{0,0.5,0}

\lstdefinestyle{PythonStyle}
{
        language=Python,
        basicstyle=\ttfamily,
        upquote=true,
        numbers = none,
        numberstyle=\footnotesize,
        escapechar=`,
        float,
        moredelim=[il][]{--latexlabel},
        otherkeywords={self},             
        commentstyle=\color{blue},
        keywordstyle=\bfseries\color{black},
        emph={MyClass,__init__,Config,HighThroughputExecutor,SlurmProvider,LocalChannel}, 
        emphstyle=\bfseries\color{deepred},    
        showstringspaces=false            %
}

\lstdefinestyle{PythonStyleInLine}
{
        language=Python,
        basicstyle=\small\ttfamily,
        upquote=true,
        numbers = none,
        numberstyle=\footnotesize,
        escapechar=`,
        moredelim=[il][]{--latexlabel},
        otherkeywords={self},             
        commentstyle=\color{blue},
        keywordstyle=\bfseries\color{black},
        emph={MyClass,__init__,@python_app,@bash_app},          
        emphstyle=\bfseries\color{deepred},    
        showstringspaces=false,            %
        frame=bt
}

\newif\iffinal

\finaltrue

\iffinal
  \newcommand{\ian}[1]{}
  \newcommand{\ryan}[1]{}
  \newcommand{\kyle}[1]{}
  \newcommand{\zhuozhao}[1]{}
  \newcommand{\yadu}[1]{}
  \newcommand{\tyler}[1]{}
  \newcommand{\ben}[1]{}

\else
  \newcommand{\ian}[1]{{\textcolor{red}{ Ian: #1 }}}
  \newcommand{\ryan}[1]{{\textcolor{magenta}{ Ryan: #1 }}}
  \newcommand{\kyle}[1]{{\textcolor{purple}{ Kyle: #1 }}}
  \newcommand{\zhuozhao}[1]{{\textcolor{cyan}{ Zhuozhao: #1 }}}
  \definecolor{darkgreen}{rgb}{0,0.5,0}
  \newcommand{\tyler}[1]{{\textcolor{green}{ Tyler: #1 }}}
  \newcommand{\yadu}[1]{{\textcolor{orange}{ Yadu: #1 }}}
  \definecolor{pink}{rgb}{1.0,0,0.5}
  \newcommand{\ben}[1]{{\textcolor{deepgreen}{ Ben: #1 }}}

\fi

\newif\ifchange

\ifchange
  
\else
  
\fi

\newcommand\mycode[1]{{\texttt{\small #1}}}

\author{Ryan Chard}
\authornote{Both authors contributed equally to the paper}
\affiliation{%
  \institution{Argonne National Laboratory}
}
\email{rchard@anl.gov}

\author{Tyler J. Skluzacek}
\authornotemark[1]
\affiliation{%
  \institution{University of Chicago}
}
\email{skluzacek@uchicago.edu}

\author{Zhuozhao Li}
\affiliation{%
  \institution{University of Chicago}
}
\email{zhuozhao@uchicago.edu}

\author{Yadu Babuji}
\affiliation{%
  \institution{University of Chicago}
}
\email{yadunand@uchicago.edu}

\author{Anna Woodard}
\affiliation{%
  \institution{University of Chicago}
}
\email{annawoodard@uchicago.edu}

\author{Ben Blaiszik}
\affiliation{%
  \institution{University of Chicago}
}
\email{blaiszik@uchicago.edu}

\author{Steven Tuecke}
\affiliation{%
  \institution{University of Chicago}
}
\email{tuecke@uchicago.edu}

\author{Ian Foster}
\affiliation{%
  \institution{Argonne \& University of Chicago}
}
\email{foster@anl.gov}

\author{Kyle Chard}
\affiliation{%
  \institution{University of Chicago}
}
\email{chard@uchicago.edu}

\renewcommand{\shortauthors}{R. Chard et al.}

\setcopyright{none}
\renewcommand\footnotetextcopyrightpermission[1]{}
\begin{document}
\title[Serverless Supercomputing]{Serverless Supercomputing: \\High Performance Function as a Service for Science}

\begin{abstract}
Growing data volumes and velocities are driving exciting 
new methods across the sciences in which data analytics and 
machine learning are increasingly intertwined with research. 
These new methods require new approaches for scientific 
computing in which computation is mobile, so that, for example, 
it can occur near data, be triggered by events (e.g., arrival of 
new data), or be offloaded to specialized accelerators. 
They also require new design approaches in which monolithic applications 
can be decomposed into smaller components, that may in turn
be executed separately and on the most efficient resources. 
To address these needs we propose \name{}---a high-performance 
function-as-a-service (FaaS) platform that enables intuitive, flexible, 
efficient, scalable, and performant remote function execution
on existing infrastructure including clouds, clusters, and supercomputers. It allows users to 
register and then execute Python functions without 
regard for the physical resource location, scheduler architecture, or 
virtualization technology on which the function is executed---an approach
we refer to as ``serverless supercomputing.''
We motivate the need for \name{} in science, describe 
our prototype implementation, and demonstrate, via experiments on two supercomputers, 
that \name{} can process millions of functions across 
more than \num{65000} concurrent workers.  We also outline
five scientific scenarios in which \name{} has been deployed
and highlight the benefits of \name{} in these scenarios.
\end{abstract}

\maketitle

\section{Introduction}

The idea that one should be able to compute wherever makes the most sense---wherever a suitable computer
is available, software is installed, or data are located, for example---is far from new: indeed, it predates the Internet~\cite{fano1965mac,parkhill1966challenge}, and motivated initiatives such as grid~\cite{Foster2001} and peer-to-peer computing~\cite{milojicic2002peer}. 
But in practice remote computing has long been complex and expensive, due to, for example, 
slow and unreliable network communications, security challenges, and heterogeneous computer architectures.

Now, however, with quasi-ubiquitous high-speed communications, universal trust fabrics, and containerization,
computation can occur essentially anywhere:
for example, where data or specialized software are located, or where computing is fast, plentiful, and/or cheap.
Commercial cloud services have embraced this new reality~\cite{varhese19cloud},
in particular via their
function as a service (FaaS)~\cite{baldini2017serverless,fox2017status} offerings 
that make invoking remote functions trivial.
Thus one simply writes \mycode{client.invoke(FunctionName="F", Payload=D)} to invoke a remote 
function \mycode{F(D)} on the AWS cloud from a Python program. 
These developments are transforming how computing is deployed and applied.
For example, Netflix uses Amazon Lambda to encode thousands of small video chunks,
make data archiving decisions, 
and validate that cloud instances adhere to security policies~\cite{netflixlambda}.
In effect, they transformed a monolithic application into one that uses event-based triggers
to dispatch tasks to where data are located, or where execution is more efficient and reliable.  

There is growing awareness of the benefits of FaaS in science and 
engineering~\cite{foster2017cloud,spillner2017faaster,malawski2016towards,fox2017conceptualizing,kiar2019serverless}, as researchers realize that their applications, too, can
benefit from decomposing monolithic applications into functions that 
can be more efficiently executed on remote computers, the use of
specialized hardware and/or software that is only available on remote computers, 
moving data to compute and vice versa, and the ability to respond to event-based triggers for computation. 
Increasingly, scientists are aware of the need for computational fluidity. 
For example, physicists at FermiLab report that a data analysis task 
that takes two seconds on a CPU can be dispatched to an FPGA device on 
the AWS cloud, where it takes 30 msec to execute, for a total of 50 msec 
once a round-trip latency of 20 msec to Virginia is included: a speedup of 40$\times$~\cite{duarte2018fast}.
Such examples arise in many scientific domains. However, 
until now, managing such fluid computations has required herculean efforts
to develop customized infrastructure to allow such offloading.  

In many ways research cyberinfrastructure (CI) is lagging with respect to 
the perpetually evolving requirements of scientific computing. 
We observe a collection of crucial challenges that lead to a significant 
impedance mismatch between sporadic research workloads and research CI
including the technical gulf between batch jobs and function-based workloads, 
inflexible authentication and authorization models, and unpredictable 
scheduling delays for provisioning resources, to name just a few. 
We are motivated therefore by the need to overcome these challenges
and enable computation of short-duration tasks (i.e., at the level 
of programming functions) with low latency and at scale across a diverse 
range of existing infrastructure, including clouds, clusters, and supercomputers. 
Such needs arise when executing machine learning inference tasks~\cite{ishakian2018}, 
processing data streams generated by instruments~\cite{malawski2016towards}, 
running data transformation and manipulation tasks on edge devices~\cite{papageorgiou2015real},
or dispatching expensive computations from edge devices to more capable systems elsewhere in the computing continuum. 

In response to these challenges we have developed a flexible, scalable, 
and high-performance function execution platform, \name{}, that adapts the powerful and flexible FaaS
model to support science workloads, and in particular data and learning system workloads, across diverse research CI.

\name{} leverages modern programming practices to
allow researchers to register functions (implemented in Python) 
and then invoke those functions on supplied input JSON documents.
\name{} manages the deployment and execution of those functions on remote resources, 
provisioning resources, staging function code and input documents, managing safe and secure
execution sandboxes using containers, monitoring execution, and returning output documents to users. 
Functions are able to execute on any compute resource where 
\name{} endpoint software is installed
and a requesting user is authorized to access. 
\name{} agents can turn \emph{any} existing resource (e.g., cloud, cluster, supercomputer, 
or container orchestration cluster) into a FaaS endpoint.

The contributions of our work are as follows: 
\begin{itemize}
	\item A survey of commercial and academic FaaS platforms and a discussion 
    of their suitability for science use cases on HPC.
    \item A FaaS platform that can: be deployed on research CI, handle
          dynamic resource provisioning and management, use various
          container technologies, and facilitate secure, scalable, and federated function execution.
    \item Design and evaluation of performance enhancements for function
          serving on research CI, including memoization, function warming, batching, and prefetching.
    \item Experimental studies showing that \name{} delivers execution latencies comparable to those
    of commercial FaaS platforms 
          and scales to 1M+ functions across 65K active workers on two supercomputers.
    \item Description of five scientific use cases that make use of \name{}, and analysis of what these use cases reveal concerning  
          the advantages and disadvantages of FaaS.
\end{itemize}

The remainder of this paper is organized as follows. 
\S\ref{sec:survey} presents a brief survey of FaaS platforms.
\S\ref{sec:background} outlines three systems built upon by \name{}.
\S\ref{sec:funcx} presents a conceptual model of \name{}.
\S\ref{sec:arch} describes the \name{} system architecture. \S\ref{sec:evaluation} and \S\ref{sec:usecases} evaluate the performance of \name{} and present five scientific case studies, respectively. Finally, \S\ref{sec:conclusion} summarizes our contributions. 

\section{A Brief Survey of FaaS}\label{sec:survey}

FaaS platforms have proved wildly successful in industry as a way
to reduce costs and the need to manage infrastructure.
Here we present a brief survey of FaaS platforms, summarized in \tablename{~\ref{tab:survey}}. 
We broadly categorize platforms as \emph{commercial}, \emph{open source}, or \emph{academic}, 
and further compare them based on the following categories.

\begin{itemize}
 \item \textbf{Languages:} The programming languages that can be used to define functions.
 \item \textbf{Infrastructure:} Where the FaaS platform is deployed and where functions are executed, e.g., cloud, Kubernetes.
 \item \textbf{Virtualization:} The virtualization technology used to isolate and deploy functions.
 \item \textbf{Triggers:} How functions are invoked and whether specific event sources are supported.
 \item \textbf{Walltime:} How long functions are permitted to execute.
 \item \textbf{Billing:} What billing models are used to recoup costs.
\end{itemize}

\definecolor{Gray}{gray}{0.9}
\newcolumntype{a}{>{\columncolor{Gray}}c}

\begin{table*}[]
\caption{Taxonomic survey of common FaaS platforms.} \label{tab:survey}

\vspace{-1ex}

\footnotesize
\begin{tabular}{|p{1.8cm}|p{2.7cm}|p{2.5cm}|p{2cm}|p{1.7cm}|p{1.6cm}|p{2cm}|}
\hline
\textbf{} & \textbf{Function Language} & \textbf{Intended Infrastructure} & \textbf{Virtualization} & \textbf{Triggers} & \textbf{Maximum Walltime (s)} & \textbf{Billing} \\ \hline
\textbf{Amazon Lambda} & C\#, Go, Java, Powershell, Ruby, Python, Node.js & Public cloud, Edge (Greengrass) & Firecracker (KVM) &  HTTP, AWS services & 900 & Requests, runtime, memory \\ \hline
\textbf{Google Cloud Functions} & BASH, Go, Node.js, Python & Public cloud & Undefined & HTTP, Pub/Sub, storage & 540 &Requests, runtime, memory \\ \hline
\textbf{Azure \newline Functions} & C\#, F\#, Java, Python, JavaScript & Public cloud, local & OS images & HTTP, APIM, MS services &600 & Requests, runtime, SLA \\ \hline
\textbf{OpenWhisk} & Ballerina, Go, Java, Node.js, Python & Kubernetes, Private cloud, Public cloud & Docker & HTTP, IBM Cloud, OW-CLI & 300 & IBM Cloud: Requests, runtime Local: NA \\ \hline
\textbf{Kubeless} & Node.js, Python .NET, Ruby Ballerina, PHP & Kubernetes & Docker & HTTP, scheduled, Pub/Sub & Undefined & NA \\ \hline 
\textbf{SAND} & C, Go, Java, Node.js, Python & Public cloud, Private cloud & Docker & HTTP, Internal event & Undefined & Triggers \\ \hline
\textbf{Fn} & Go, Java, Ruby, Node.js, Python &  Public cloud, Kubernetes & Docker & HTTP, direct trigger & 300 & NA \\ \hline
\textbf{Abaco} & Container & TACC clusters & Docker &  HTTP & Undefined & Undefined \\ \hline
\rowcolor[HTML]{C0C0C0} 
\textbf{\name{}} & Python & Local, clouds, clusters, supercomputers & Singularity, Shifter, Docker & HTTP, Globus Automate & No limit & HPC SUs, cloud credits. Local: NA \\ \hline
\end{tabular}
\end{table*}

\subsection{Commercial Platforms}
Most commercial cloud providers offer FaaS capabilities.
Here we compare three 
platforms offered by Amazon, Microsoft, and Google. 

\emph{Amazon Lambda}~\cite{AmazonLambda} pioneered the FaaS paradigm in 2014 and has since
be used in many 
industry~\cite{netflixlambda} and academic~\cite{chard17ripple} use cases. Lambda is a hosted service
that supports a multitude of function languages and trigger sources (Web interface, CLI, SDK, and other AWS services). 
Tight integration with the wider AWS ecosystem means Lambda functions can be associated with triggers from other AWS services, such as CloudWatch, S3, API gateways, SQS queues, and Step Functions. Functions are billed based on their memory allocation and 
for every 100ms execution time. Once defined, Lambda uses a custom virtualization technology built on KVM, called Firecracker to create lightweight micro-virtual machines. These microVMs then persist in a \textit{warmed} state for five minutes and continue to serve requests. 
While Lambda is provided as a hosted service, functions can be deployed locally or to edge devices 
via the Greengrass~\cite{greengrass} IoT platform. 

\emph{Google Cloud Functions}~\cite{googlecloudfunctions} is differentiated by its tight coupling to Google Cloud Storage, Firebase mobile backends, and custom IoT configurations via Google's globally distributed message bus (Cloud Pub/Sub).
Like Lambda, Google Cloud Functions also support triggers from arbitrary HTTP webhooks. 
Further, users can trigger functions through a number of third party systems including 
GitHub, Slack, and Stripe. 
While Google Cloud functions apply a similar pricing model to Lambda, the model is slightly more expensive
for high-volume, less computationally intensive tasks as Lambda has lower per-request costs after the first two million invocations (with similar compute duration costs). 

\emph{Azure Functions}~\cite{azureFunctions} allow users to create functions in a native language through either the Web interface or the CLI. Functions are packaged and may be tested locally using a local web service before being uploaded to the Azure platform.
Azure functions integrate with other Azure products through triggers. 
Triggers are provided from CosmosDB, Blob storage, and Azure storage queues, in addition to custom HTTP and time-based triggers. Azure price-matches AWS for compute and storage (as of November 2018).

\subsection{Open Source Platforms}
Open FaaS platforms resolve two of the key challenges to using FaaS for scientific workloads: they 
can be deployed on-premise and can be customized to meet the requirements of data-intensive 
workloads without any pricing models.

\emph{Apache OpenWhisk}~\cite{openwhisk} is the most well-known open source FaaS platform. OpenWhisk is the basis of IBM Cloud Functions~\cite{IBMCloudFunctions}. OpenWhisk clearly defines an event-based programming model, consisting of \emph{Actions} which are stateless, runnable functions, \emph{Triggers} which are the types of events OpenWhisk may track, and \emph{Rules} which associate one trigger with one action. OpenWhisk can be deployed locally as a service using a Kubernetes cluster. However, deploying OpenWhisk is non-trivial, requiring installation of dependencies and administrator access to the cluster. 

\emph{Fn}~\cite{Fn} is a powerful open-source software from Oracle that can be deployed on any Linux-based compute resource having administrator access to run Docker containers. Applications---or groups of functions---allow users to logically group functions to build runtime utilities (e.g.,
dependency downloads in custom Docker containers) and other resources (e.g.,
a trained machine learning model file) to support functions in the group. Moreover, Fn supports fine-grained logging and metrics, and is one of few open source FaaS platforms deployable on Windows. Fn can be deployed locally or on a Kubernetes cluster. In our experience, one can deploy a fully-functional Fn server in minutes.

\emph{Kubeless}~\cite{Kubeless} is a native Kubernetes FaaS platform that takes advantage of built-in Kubernetes primitives. Kubeless uses Apache Kafka for messaging, provides a CLI that mirrors that of AWS Lambda, and supports fine-grained monitoring. Users can invoke functions via the CLI, HTTP, and via a Pub/Sub mechanism. Like Fn, Kubeless allows users to define function groups that share resources.  Like OpenWhisk, Kubeless is reliant on Kubernetes and cannot be deployed on other resources.

\subsection{Academic Platforms}

The success of FaaS in industry has spurred academic exploration of FaaS. Two
systems that have resulted from that work are SAND~\cite{akkus2018sand} and Actor Based Co(mputing)ntainers (Abaco)~\cite{stubbs2017containers}.

\emph{SAND}~\cite{akkus2018sand} is a lightweight, low-latency FaaS platform from Nokia Labs that provides application-level sandboxing and a hierarchical message bus. The authors state that they achieve a 43\% speedup and a 22x latency reduction over Apache OpenWhisk 
in commonly-used image processing applications.
Further, SAND provides support for function or \textit{grain} chaining via user-submitted workflows. At the time of their writing, it appears that SAND does not support multi-tenancy, only having isolation at the application level.
SAND is closed source and as far as we know cannot be downloaded and installed locally.

\emph{Abaco}~\cite{stubbs2017containers} supports functions written in a wide range of programming languages and supports automatic scaling. Abaco implements the Actor model in which an \textit{actor} is an Abaco runtime mapped to a specific Docker image. Each actor executes in response to messages posted to its \textit{inbox}. Moreover, Abaco provides fine-grained monitoring of container, state, and execution events and statistics. Abaco is deployable via Docker Compose.

\subsection{Summary of FaaS}
Commercial cloud providers implement high performance and reliable FaaS models 
that are used by huge numbers of users. However, for science use cases they 
are unable to make use of existing infrastructure, they do not integrate
with the science ecosystem (e.g., in terms of data and authentication models), 
and they can be costly. 

Open source and academic frameworks support on-premise deployments and can be 
configured to address a range of use cases. However, each of the systems surveyed
is Docker-based and therefore requires administrator privileges to be deployed 
on external systems. Furthermore, the reliance on Docker prohibits use in 
most computing centers which instead support user space containers. 
In most cases, these systems have been implemented to rely on Kubernetes (or other
container orchestration models such as Mesos and Openshift) which means
they cannot be adapted to existing HPC and HTC environments. 

\name{} provides a scalable, low-latency 
FaaS platform that can be applied to existing HPC resources with minimal effort. It employs user-space
containers to isolate and execute functions, avoiding the security concerns prohibiting other FaaS
platforms from being used. Finally, it provides an intuitive interface for executing scientific
workloads and includes a number of performance optimizations to support broad scientific use cases.

\subsection{Other Related Approaches}
FaaS builds upon a large amount of related work including
in Grid and cloud computing, container orchestration, and analysis systems. 
Grid computing~\cite{Foster2001} laid the foundation for 
remote, federated computations, most often applying 
federated batch submission~\cite{krauter02gridmanagement}. 
GridRPC~\cite{seymour02gridrpc} defines an API for executing 
functions on remote servers requiring that developers implement
the client and the server code. 
\name{} extends these ideas to allow interpreted functions
to be registered and subsequently to be dynamically 
executed within sandboxed containers via a standard endpoint API.

Container orchestration systems~\cite{rodriguez19containers, hightower17kubernetes, hindman11mesos} allow
users to scale deployment of containers while 
managing scheduling, fault tolerance, resource provisioning, and addressing
other user requirements. These systems primarily rely on dedicated, cloud-like
infrastructure and cannot be directly applied to HPC resources. \name{} 
provides similar functionality, 
however it focuses at the level of scheduling and managing functions, that are deployed
across a pool of containers. We apply approaches from container orchestration
systems (e.g., warming) to improve performance.

Data-parallel systems such as Hadoop~\cite{hadoop} and Spark~\cite{spark}
enable map-reduce style analyses. 
Unlike \name{}, these systems dictate a particular programming
model on dedicated clusters. 
Parallel computing libraries such as Dask~\cite{dask}, Parsl~\cite{babuji19parsl}, 
and Ray~\cite{moritz2018ray} support parallel execution of scripts, 
and selected functions within those scripts, 
on clusters and clouds. \name{} uses Parsl to manage
function execution in containers.

\section{Background}
\label{sec:background}

We build \name{} on a foundation of existing work, including
the Parsl parallel scripting library~\cite{babuji19parsl} 
and Globus~\cite{chard2014efficient}. 

\subsection{Parsl}

Parsl is parallel scripting library that augments Python with simple, scalable, and flexible constructs for encoding parallelism. Parsl is designed for scalable execution of Python-based workflows on a variety of resources---from laptops to clouds and supercomputers. It includes an 
extensible set of executors tailored to different use cases, such as low-latency, high-throughput, or extreme-scale execution. Parsl's modular executor architecture enables
users to port scripts between different resources, and scale from small clusters
through to the largest supercomputers with many thousands of nodes and tens of thousands of workers. Here we use Parsl's high-throughput executor as the base for the 
\name{} endpoint software as it provides scalable and reliable execution of functions.

Parsl is designed to execute workloads on various resource types, such as AWS, Google Cloud, Slurm, PBS, Condor, and many others. 
To do so, it defines a common \emph{provider} interface that can acquire 
(e.g., via a submission script or cloud API call), monitor, and manage
resources. Parsl relies on a Python configuration object to define and
configure the provider. \name{} uses Parsl to connect
to various resources and adopts Parsl's configuration object to define
how a deployed endpoint should use its local resources.

\subsection{Globus}

Globus Auth~\cite{GlobusAuth} provides authentication and authorization platform
services designed to support an ecosystem of services, applications, and clients
for the research community. 
It allows external services (e.g., the \name{} service and the \name{} endpoints)
to outsource authentication processes such that users may authenticate using
one of more than 600 supported
identity providers (e.g., Google, ORCID, and campus credentials).
Services can also be registered as Globus Auth resource servers, 
each with one or more unique \emph{scopes} (e.g., \emph{execute\_function}).
Other applications and services may then
obtain delegated access tokens (after consent by a user or client) to securely
access other services as that user (e.g., to register or invoke a function). 
We rely on Globus Auth throughout the \name{} architecture and in particular
to provide user/client authentication with the system and to support secure 
endpoint registration and operations with the \name{} service. 

\section{Conceptual Model}\label{sec:funcx}

We first describe the conceptual model behind \name{},
to provide context to the implementation architecture. 
\name{} allows users to register and then execute \emph{functions} 
on arbitrary \emph{endpoints}.
All user interactions with \name{} are performed via a 
REST API implemented by a cloud-hosted \name{} service. 
Interactions between users, the \name{} service, and 
endpoints are subject to Globus Auth-based authentication
and authorization.

\textbf{Functions:}
\name{} is designed to execute \emph{functions}---snippets of Python code
that perform some activity. A \name{} function explicitly defines
a function body that contains the entire function, takes a JSON 
object as input and may return a JSON object. The function
body must specify all imported modules. 
Functions must be registered before they can be invoked by
the registrant or, if permitted, other users.
An example function for processing raw tomographic data is shown in Listing~\ref{lst:function}. 
This function is used create a tomographic preview image from a
HDF5 input file. The function's input specifies the file and 
parameters to identify and read a projection. 
It uses the automo Python package to read the data, 
normalize the projection, and then save the preview image. 
The function returns the name of the saved preview image.

\begin{lstlisting}[style=PythonStyleInLine,caption={Python function to create neurocartography preview images from tomographic data. \label{lst:function}}]
def automo_preview(event):
  import numpy, tomopy
  from automo.util import (read_data_adaptive, 
                           save_png
  data = event['data']
  proj, flat, dark, _ = read_data_adaptive(
        data['fname'], proj=(data['st'], 
        data['end'], data['step'])
      )    
  proj_norm = tomopy.normalize(proj, flat, dark)
  flat = flat.astype('float16')
  save_png(flat.mean(axis=0), fname=('prev.png'))

  return {'filename': 'prev.png'}
\end{lstlisting}

\textbf{Endpoints:}
A \name{} endpoint is a logical interface to a computational resource
that allows the \name{} service to dispatch function invocations to that resource.
The endpoint handles authentication and authorization, 
provisioning of nodes on the compute resource, 
and various monitoring and management functions.
Users can download the \name{} endpoint software, deploy it on a
target resource, and register it with \name{}
by supplying connection information and metadata (e.g., name and description). 
Each registered endpoint is assigned a unique identifier for subsequent use.

\textbf{Function execution:}
Authorized users may invoke a registered function on a selected endpoint. 
To do so, they issue a request via the \name{} service which identifies 
the function and endpoint to be used as well as an input JSON document
to be passed to the function.  Optionally, the user may
specify a container image to be used. This 
allows users to construct environments with appropriate dependencies
(system packages and Python libraries) required to execute the function. 
Functions may be executed synchronously or asynchronously;
in the latter case the invocation returns an
identifier via which progress may be monitored and results retrieved. 

\textbf{Web service:}
The \name{} service exposes a REST API for registering functions and 
endpoints, and for executing functions, managing their execution, and retrieving results. 
The Web service is paired with accessible endpoints via the endpoint
registration process. 
The \name{} service is a Globus Auth resource server and thus enables
users to login using an external identity and for programmatic access
via OAuth access tokens.

\textbf{User interface:}
\name{} is designed to be used via the REST API or \name{} Python SDK 
that wraps the REST API. 
Listing~\ref{lst:sdk} shows an example of how the SDK can be used
to invoke a registered function on a specific endpoint.
The example first imports the \texttt{FuncXClient}, it then
constructs a client, defaulting to the address of the
public \name{} web service. 
It then invokes a registered function using the \texttt{run}
command and passes the unique function identifier, a JSON
document with input data (in this case the path to a file), 
the endpoint id on which to execute the function,  
the \emph{funcx\_python3.6} container in which the function
will be executed, and it also sets the interaction to be asynchronous. 
Finally, the example shows that the function can be monitored
using \texttt{status} and the asynchronous results retrieved
using \texttt{result}.

\begin{lstlisting}[style=PythonStyleInLine, caption={Example use of the \name{} SDK to invoke a registered function in the funcx\_python3.6 container. \label{lst:sdk}}]
from funcx import FuncXClient
fx = FuncXClient()

func_id = '6d79-...-764bb'
container_name = 'funcx_python3.6'
endpoint_id = '863d-...-d820d'
data = {'input': '/projects/funcX/test.h5'}
func_res = fx.run(func_id, data, endpoint_id, 
                  container_name, async=True)
func_res.status()
func_res.result()
\end{lstlisting}

\section{Architecture and Implementation}\label{sec:arch}
\begin{figure}[h]
  \includegraphics[width=0.9\columnwidth]{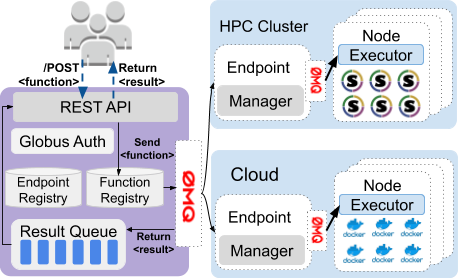}
  \vspace{-0.1in}
  \caption{\name{} architecture showing the \name{} service on the left and
two \name{} endpoints deployed on a Cloud and HPC cluster on the right. Each endpoint's manager is responsible for coordinating execution of functions
via executors deployed on nodes.
\label{fig:arch}
\vspace{-0.1in}
}
\end{figure}

The \name{} system combines a cloud-hosted management service with 
software agents---\name{} endpoints---deployed on remote resources.  
The cloud-hosted \name{} service implements endpoint management and 
function registration, execution and management. 
\name{}'s primary interface is the hosted REST API; 
a Python SDK supports use in programming environments
and integration in external applications.
The advantages of such a service-oriented model are well-known and include 
ease of use, availability, reliability, and reduced software 
development and maintenance costs. 
An overview of \name{}'s architecture
is depicted in \figurename{~\ref{fig:arch}}.

\subsection{The \name{} Service}
The \name{} service maintains a registry of \name{} endpoints and registered
functions. The service provides a REST API to register and manage endpoints, 
register functions, and execute, monitor, and retrieve the output from functions. 
The \name{} service is secured using Globus Auth allowing users
to authenticate with it directly (e.g., via the native app flow
in a Jupyter notebook) or via external clients that 
can call the REST API directly. 
It also allows for endpoints, registered as Globus Auth clients, 
to call the API to register themselves with the service.
The \name{} service is implemented in Python as a Flask application, 
it is deployed on AWS
and relies on Amazon Relation Database Service (RDS) to store registered endpoints and functions.

\subsection{Function Containers} 
\name{} uses containers to package function code that is to be deployed
on a compute resource. 
Key requirements for a packaging technology include portability 
(i.e., a package can be deployed in many different environments with little or no change)
completeness (all code and dependencies required to run a function can be captured), 
performance (minimal startup and execution overhead; small storage size),
and safety (unwanted interactions between function and environment can be avoided).
Container technology meets these needs well.

Our review of container technologies, including Docker~\cite{merkel2014docker}, LXC~\cite{LXC}, Singularity~\cite{kurtzer2017singularity}, Shifter~\cite{jacobsen2015contain}, and CharlieCloud~\cite{priedhorsky2017charliecloud}, 
leads us to adopt Docker, Singularity, and Shifter in the first instance.
Docker works well for local and cloud deployments, whereas
Singularity and Shifter are designed for use in HPC environments
and are supported at large-scale computing facilities (e.g., Singularity at ALCF and Shifter at NERSC). 
Singularity and Shifter implement similar models and thus 
it is easy to convert from a common representation (i.e., a Dockerfile)
to both formats. 

\name{} requires
that each container includes a base set of software, including Python~3 
and \name{} worker software.  In addition, any other system libraries or Python 
modules needed for function execution must be added manually to the container. 
When invoking a function, users must specify the container to be used for execution;
if no container is specified, \name{} uses a base \name{} image. 
In future work, we intend to make this process dynamic, using repo2docker~\cite{repo2docker}
to build Docker images and convert them to site-specific container formats
when needed.

\subsection{The \name{} Endpoint}\label{sec:executor}

The \name{} endpoint represents the remote computational resource
(e.g., cloud, cluster, or supercomputer) upon which it is deployed.
The endpoint is designed to deliver high-performance execution of functions in a secure, scalable and reliable manner.

The endpoint architecture, depicted in \figurename{~\ref{fig:executor}, is comprised of three components, which are discussed below:
\begin{itemize}
\item \emph{Manager}: queues and forwards function execution requests and results, 
interacts with resource schedulers, and batches and load balances requests.
\item \emph{Executor}: creates and manages a pool of workers on a node.
\item \emph{Worker}: executes functions within a container.
\end{itemize}

The \emph{Manager} is the daemon that is deployed by a user on a 
HPC system (often on a login node) 
or on a dedicated cloud node.
It authenticates with the \name{} service and upon registration acts as a conduit for routing functions and results
between the service and workers. 
A manager is responsible for managing resources on its system
by working with the local scheduler or cloud API to deploy 
executors on compute nodes.
The manager uses a pilot job model~\cite{turilli2018comprehensive} to connect to
and manage resources in a uniform manner,
irrespective of the resource type (cloud or cluster) or local resource manager (e.g., Slurm, PBS, Cobalt).
As each executor is launched on a compute node, it connects to and registers with the manager. 
The manager then uses ZeroMQ sockets to communicate 
with its executors. 
To minimize blocking, all communication is managed by threads
using asynchronous communication patterns. 
The manager uses a randomized scheduling algorithm to allocate functions to executors. 

To provide fault tolerance and robustness, for example with respect to node failures, the manager uses heartbeats and a watchdog
process to detect failures or lost executors. The manager tracks tasks that have been distributed to executors so that when failures
do occur, lost tasks can be re-executed (if permitted). 
Communication from \name{} service to managers uses the reliable Majordomo
broker pattern in ZeroMQ. Loss of a manager is terminal and relayed to the user.
To reduce overheads, the manager can shut down executors when they are not
needed; suspend executors to prevent further tasks being scheduled to failed executors;
and monitor resource capacity to aid scaling decisions.

\emph{Executors} represent, and communicate on behalf of, the collective capacity of the workers on a single node, thereby limiting
the number of sockets used to just two per node. Executors determine the available CPU/Memory resources on a node, and partition the
node amongst the workers. Once all workers connect to the executor, it registers itself with the manager. Executors advertise available
capacity to the manager, which enables batching on the executor.

\emph{Workers} persist within containers and each executes one function at a time. Since workers have a single responsibility
they use blocking communication to wait for functions from the executor. Once a function is received it is deserialized and executed,
and the serialized results are returned via the executor.

\subsection{Managing Compute Infrastructure}
The target computational resources for \name{} range from 
local deployment to clusters, clouds, and supercomputers 
each with distinct modes of access.
As \name{} workloads are often sporadic, resources must be provisioned 
as needed so as to reduce startup overhead and wasted allocations.
\name{} uses Parsl's provider interface~\cite{babuji19parsl} to 
interact with various resources, specify resource-specific
requirements (e.g., allocations, queues, limits, or cloud instance types), 
and define the rules for automatic scaling (i.e., limits and scaling aggressiveness).
With this interface, \name{} can be deployed on batch schedulers 
such as Slurm, Torque, Cobalt, SGE and Condor as well as
the major cloud vendors such as AWS, Azure, and Google Cloud.

\begin{figure}[h]
\vspace{-0.1in}
  \includegraphics[width=0.9\columnwidth]{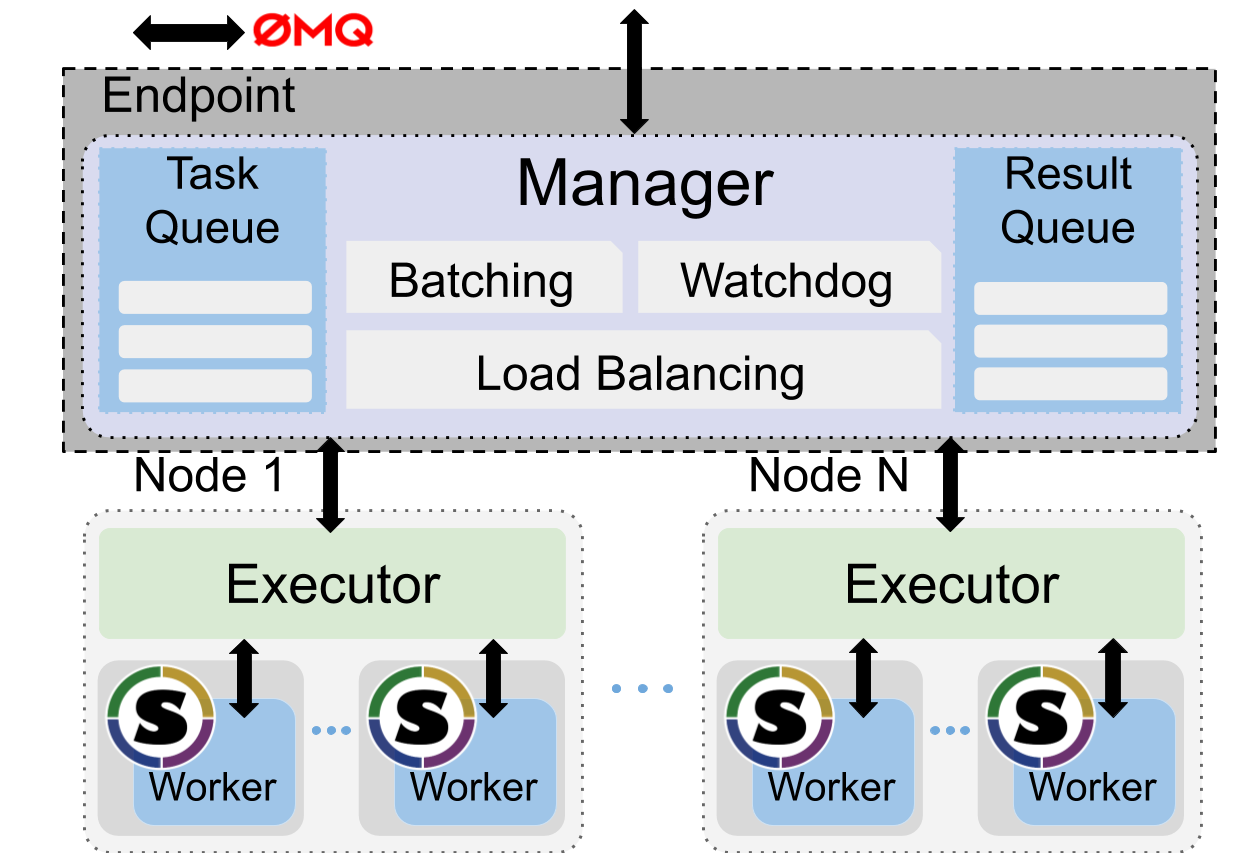}
  \vspace{-0.1in}
  \caption{The \name{} endpoint.
\label{fig:executor}
\vspace{-0.1in}
}
\end{figure}

\subsection{Optimizations}

We apply several optimizations to enable high-performance function serving in a wide range of computational environments. We briefly describe five optimization methods employed in \name{}.

\textbf{Memoization} involves returning a cached result
when the input document and function body have been processed previously. 
\name{} supports memoization by hashing the function body and input document and storing a mapping
from hash to computed results. Memoization is only used if explicitly set by the user.

\textbf{Container warming} is used by cloud FaaS platforms to improve performance~\cite{wang2018peeking}. 
Function containers are kept \emph{warm} by leaving them running for a short period of time (5-10 minutes) 
following the execution of a function. 
This is in contrast to terminating containers at 
the completion of a function. Warm containers remove the need to instantiate a new container to execute a function, 
significantly reducing latency. 
This need is especially evident in HPC resources for several reasons: first, loading many 
concurrent Python environments and containers puts a strain on large, shared file systems; 
second, many HPC centers have their own methods for instantiating containers that may place
limitations on the number of concurrent requests; and third, individual cores are often 
slower in many core architectures like Xeon Phis. As a result the start time for containers 
can be much larger than what would be seen locally.

\textbf{Batching} requests enables \name{} to amortize costs across many function requests. 
\name{} implements two batching models:
first, batching to enable executors to request many tasks on behalf of their workers, minimizing network communication costs; 
second, user-driven batching of function inputs, 
allowing the user to manage the tradeoff between more
efficient execution and increased per-function latency by choosing to create fewer, larger
requests. Both techniques can increase overall throughput.

\textbf{Prefetching} is a technique for requesting more tasks than
can be satisfied immediately in the anticipation of availability in the near future.
\name{} executors use prefetching to improve performance by requesting tasks 
while workers are busy with execution, thus interleaving network communication
with computation. This can improve performance for short, latency-sensitive functions.

\textbf{Asynchronous messaging} is a technique for hiding network latencies. 
\name{} uses asynchronous messaging patterns provided by
ZeroMQ to implement end-to-end socket based inter-process communication. 
By avoiding blocking communication patterns, \name{} ensures that even when
components over widely varying networks are connected, 
performance will not be bottlenecked by the slowest connection.

\subsection{Automation}
FaaS is often used for automated processing in response to various
events (e.g., data acquired from an instrument).
To facilitate event-based execution in research scenarios we have integrated \name{}
with the Globus Automate platform~\cite{ananthakrishnan18publication}. 
To do so we have implemented the \emph{ActionProvider} interface in \name{}
by creating REST endpoints to start, cancel, release, and check the status of the task.
Exposing \name{} as an ActionProvider allows automation flows to execute 
functions on behalf of a user. The API uses Globus Auth to determine the identity 
of the user that owns the flow, and uses their authentication tokens to execute 
functions via the \name{} service and endpoint. When specifying the action in a
flow the user must define the function ID, input JSON document, and endpoint ID for
execution. When the flow invokes the function, the \name{} service creates 
an identifier to return to the automation platform for monitoring of that 
step of the workflow. 

\subsection{Security Model}

Secure, auditable, and safe function execution is crucial to \name{}.
We implement a comprehensive security model to ensure 
that functions are executed by authenticated and authorized users 
and that one function cannot interfere another. We rely on
two proven security-focused technologies: Globus Auth~\cite{GlobusAuth}
and containers.

\name{} uses Globus Auth for authentication, authorization, and protection of all APIs. 
The \name{} service is represented as a Globus Auth resource server, allowing users to 
authenticate using a supported Globus Auth identity (e.g., institution, Google, ORCID)
and enabling various OAuth-based authentication flows (e.g., confidential client credentials, 
native client) for different scenarios. 
It also has its own unique Globus Auth scopes
(e.g., ``urn:globus:auth:scope:--funcx.org:register\_function'')
via which other services (e.g., Globus Automate) may obtain authorizations
for programmatic access. 
\name{} endpoints are registered as Globus Auth clients, each dependent on
the \name{} scopes, which can then be used to connect to the \name{} service. 
Each endpoint is configured with a Globus Auth client\_id/secret
pair which is used for constructing REST requests.
The connection between the \name{}
service and endpoints is established using ZeroMQ. Communication addresses
are communicated as part of the registration process. Inbound
traffic from endpoints to the cloud-hosted service is limited to known IP addresses. 

All functions are executed in isolated containers to ensure that functions
cannot access data or devices outside that context. 
In HPC environments we use Singularity and Shifter.
\name{} also integrates additional sandboxing procedures to isolate
functions executing within containers, namely, creating namespaced directories within the containers 
in which to capture files that are read/written. To enable fine grained tracking of execution, 
we store execution request histories in the \name{} service
and in logs on \name{} endpoints.

\section{Evaluation}\label{sec:evaluation}

We evaluate the performance of \name{} in terms of latency, scalability, throughput, and fault tolerance. We also explore the affect of batching, memoization, and prefetching.

\subsection{Latency}

To evaluate \name{}'s latency we compare it with commercial FaaS platforms by measuring the time required for single function invocations. 
We have created and deployed the same Python function (Listing~\ref{lst:latency-function}) on Amazon Lambda, Google Cloud Functions, Microsoft Azure Functions, and \name{}. To minimize unnecessary overhead we use the same payload when invoking each function: the string ``hello-world.'' Each function simply prints and returns the string.

\vspace{-0.05in}
\begin{lstlisting}[style=PythonStyleInLine,caption={Python function to calculate latency.\label{lst:latency-function}}]
def hello_world(event):
    print(event)
    return event
\end{lstlisting}

Although each provider operates its own data centers, we attempt to standardize network latencies by placing functions in an available US East region (between South Carolina and Virginia). We deploy \name{} service and endpoint on two AWS 
EC2 instances (m5.large) in the US East region. 
We use an HTTP trigger to invoke the function on each of the FaaS platforms.
We then measure latency as the round-trip time to submit, execute, and return a result from the function. 
We submit all requests from the login node of Argonne National Laboratory's Cooley cluster, in Chicago, IL (20.5 ms latency to the \name{} service). 
The experiment configuration is shown in \figurename~\ref{fig:experiment}.

\begin{figure}[h]
  \includegraphics[width=0.9\columnwidth]{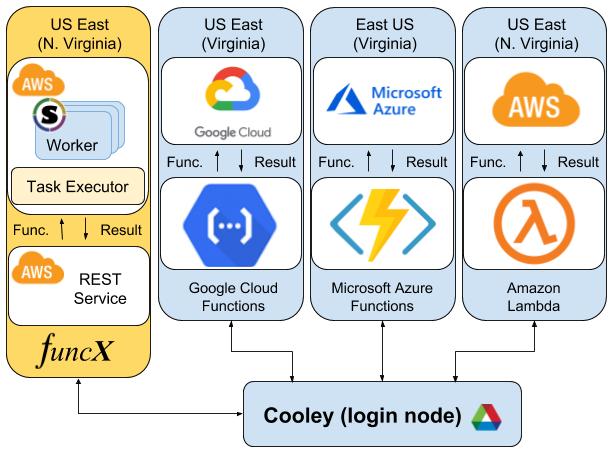}
    \vspace{-0.1in}
  \caption{Comparative latency experiment architecture.}
  \label{fig:experiment}
    \vspace{-0.1in}
\end{figure}

\begin{figure}[h]
  \includegraphics[width=0.9\columnwidth]{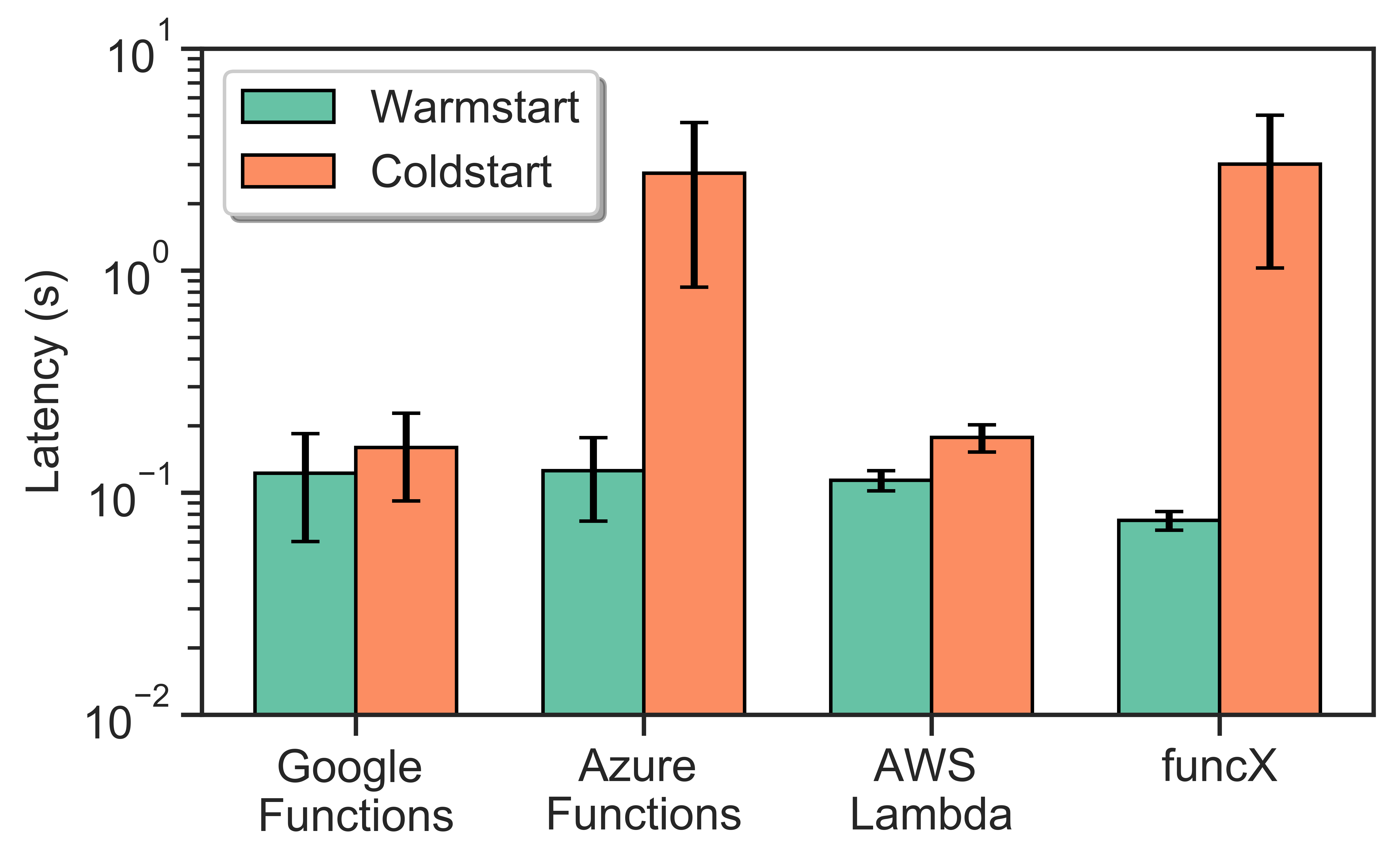}
  \vspace{-0.1in}
  \caption{Average task latency (s) over \num{2000} functions.}
  \label{fig:latency_eval}
   \vspace{-0.1in}
\end{figure}

For each FaaS service we compare the cold start time and warm start time. The cold start time
aims to capture the scenario where a function is first executed and the function code and execution environment must be
configured. To capture this in \name{} we restart the service and measure the 
time taken to launch the first function. For the other services, we simply invoke functions every 10 minutes and 1 second 
(providers report maximum cache times of 10 minutes, 5 minutes, 5 minutes, for Google, Amazon, and Azure, respectively)
in order to ensure that each function starts cold.
We execute the cold start functions 40 times, and the warmed functions \num{2000} times. 
We report the mean completion time and standard deviation for each FaaS platform in \figurename~\ref{fig:latency_eval}.
We notice that Lambda, Google Functions, and Azure Functions exhibit warmed round trip times of 116ms, 122ms, and 126ms, respectively.
\name{} proves to be considerably faster, running warm functions in 76ms. 
We suspect this is due to \name{}'s minimal overhead, as, for example, requests are sent directly to the \name{} service rather than through elastic load balancers (e.g., AWS ELB for Lambda), and also likely incur fewer logging and resiliency overheads. 
When comparing cold start performance, we find that Lambda, Google Functions, Azure Functions, and \name{} exhibit cold round trip times of 175ms, 160ms, \num{2748}ms, and \num{2886}ms respectively. 
Google and Lambda exhibit significantly lower cold start times,
perhaps as a result of the simplicity of our function (which requires only standard Python libraries
and therefore could be served on a standard container)
or perhaps due to the low overhead of these proprietary container technologies~\cite{wang2018peeking}. 
In the case of \name{} this overhead is primarily due to the startup time of the container (see Table~\ref{table:cold_start_cost}). 

We next break down the latency of each function invocation for each FaaS service. 
Table~\ref{table:lat_others} shows the total time for warm and cold functions
in terms of overhead and function execution time. 
For the closed-source, commercial FaaS systems we obtain 
function execution time from execution logs and compute overhead
as any additional time spent invoking the function. As expected, 
overheads consume much of the invocation time. Somewhat surprisingly, 
we observe that Lambda has much faster function execution time for 
cold than for warm containers, perhaps as the result of the way Amazon
reports usage. 
We further explore latency for \name{} by instrumenting the system.
The results are shown in Figure~\ref{fig:lat_breakdown} for a warm
container. Here we consider the following times: 
$t_c$: round-trip time between the \name{} client on Cooley and the \name{} service, 
$t_w$: web service latency to dispatch the request to and endpoint (and then to return the result), 
$t_m$: endpoint connection latency from receiving the request (including data transfer and queue processing) until it is passed to an executor, and 
$t_e$: function execution time.  
We observe that $t_e$ is fast relative to the overall system latency. 
$t_c$ is mostly made up of the communication time from Cooley to AWS (measured at 20.5ms).
While $t_m$ only includes minimal communication time due to AWS-AWS connections (measured at 1ms). 
Most of the \name{} overhead is therefore captured in $t_w$ as a result
of database access and endpoint routing, and $t_m$ as a result of internal queuing and 
Parsl dispatching.

\begin{figure}[h]
  \includegraphics[width=0.9\columnwidth]{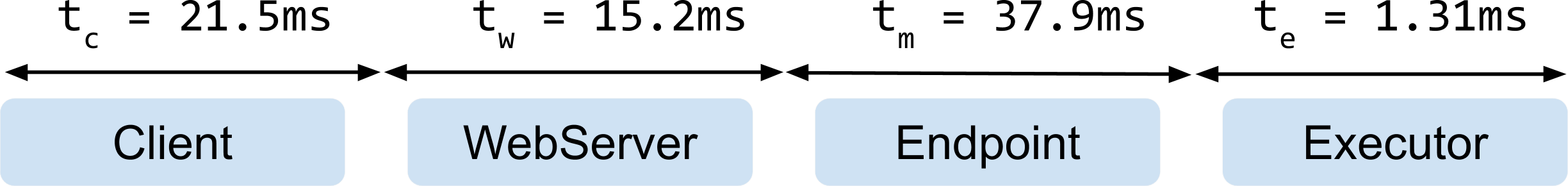}
    \vspace{-0.1in}
  \caption{\name{} latency breakdown for a warm container.}
  \label{fig:lat_breakdown}
    \vspace{-0.1in}
\end{figure}

\begin{table}[]
\footnotesize
\caption{FaaS latency breakdown (in ms).}
\vspace{-0.1in}
\begin{tabular}{l l r r r}
  \multicolumn{2}{l}{}    & \textbf{Overhead} & \textbf{Function} & \textbf{Total} \\
  \hline
  \multirow{2}*{\textbf{Azure}}      & warm &      112.0 &     13.6 &   125.6  \\
                                     & cold &    2720.0 &     28.0 & 2748.0  \\
  \multirow{2}*{\textbf{Google}}     & warm &      117.0 &      5.0 &   122.0  \\
                                     & cold &      136.0 &     24.0 &   160.0  \\
  \multirow{2}*{\textbf{Lambda}}     & warm &      116.0 &     0.3 &  116.3  \\
                                     & cold &      174.0 &     0.5 &  174.5  \\
  \multirow{2}*{\textbf{\name{}}}    & warm &       74.6 &     1.3 &   75.9  \\
                                     & cold &      2882.0 &  4.2 & 2886.0 \\
  \hline
\end{tabular}
\label{table:lat_others}
\vspace{-0.1in}
\end{table}

\subsection{Scalability and Throughput}
We study the strong and weak scaling of \name{} using  
Argonne National Laboratory's Theta~\cite{Theta} and NERSC's Cori~\cite{cori} supercomputers.
Theta is a 11.69-petaflop system based on the second-generation Intel Xeon Phi ``Knights Landing" (KNL) processor. The system is equipped with \num{4392} nodes, each containing a 64-core processor with 16 GB MCDRAM, 192 GB of DDR4 RAM, and interconnected with high speed InfiBand.
Cori consists of an Intel Xeon ``Haswell" partition and an Intel Xeon Phi KNL partition. 
Our tests were conducted on the KNL partition. 
Cori's KNL partition has \num{9688} nodes in total, 
each containing a 68-core processor (with 272 hardware threads) 
with six 16GB DIMMs, 96 GB DDR4 RAM, and interconnected with Dragonfly topology.
We perform experiments using 64 Singularity containers on each Theta node 
and 256 Shifter containers on each Cori node.
Due to a limited allocation on Cori we use the four hardware threads
per core to deploy more containers than cores.

\begin{figure}[h]
  \vspace{-0.1in}
  \includegraphics[width=\columnwidth,trim=0.07in 0.08in 0.07in 0.1in,clip]{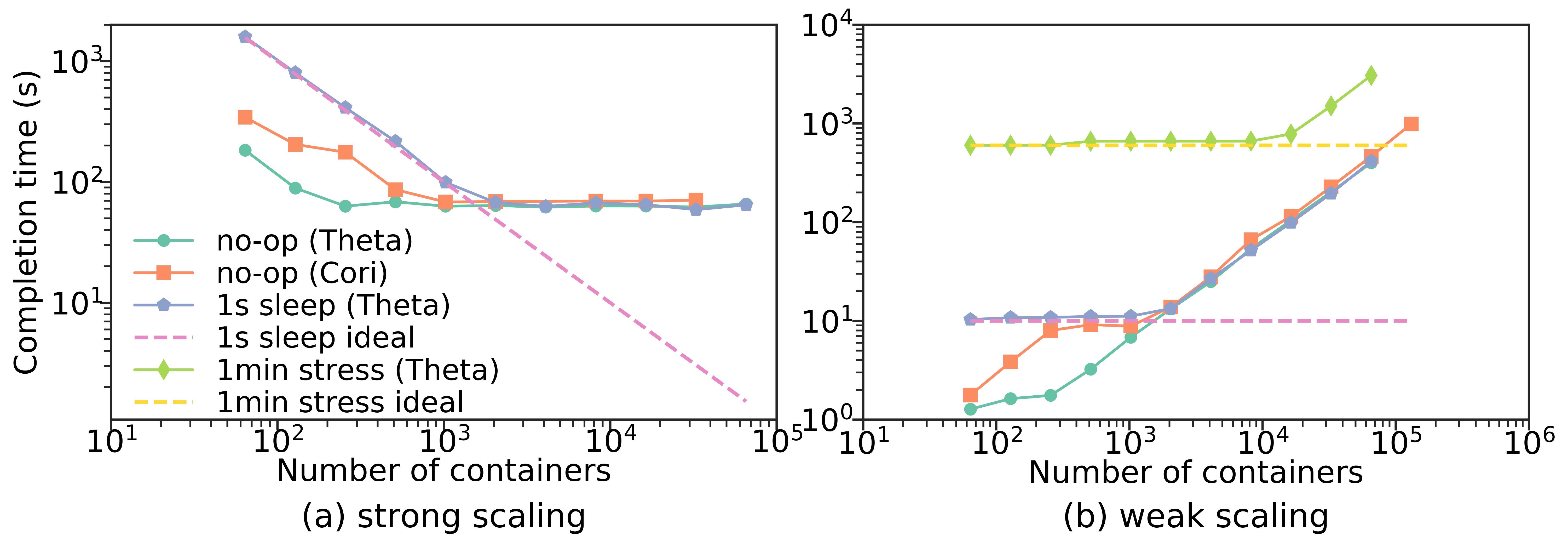}
  \vspace{-0.2in}
  \caption{Strong and weak scaling of \name{}.}
\label{fig:scalability}
  \vspace{-0.1in}
\end{figure}

Strong scaling evaluates performance when the total number of function invocations is fixed; 
weak scaling evaluates performance when the average number of functions executed on each container is fixed. 
To measure scalability we created functions of various durations:  
a 0-second ``no-op'' function that exits immediately, a 1-second ``sleep'' function, and a 1-minute CPU ``stress'' function that keeps a CPU core at 100\% utilization.
For each case, we measured completion time of a batch of functions as we increased the number of total containers.
Notice that the completion time of running $M$ ``no-op'' functions on $N$ workers indicates the overhead of \name{} 
to distribute the $M$ functions to $N$ containers.
Due to limited allocation we did not execute sleep or stress 
functions on Cori, nor did we 
execute stress functions for strong scaling on Theta.

\subsubsection{Strong scaling}

Figure~\ref{fig:scalability}(a) shows the completion time of \num{100000} 
\emph{concurrent} function requests with an increasing number of containers.
On both Theta and Cori the completion time decreases as the number of containers 
increases until we reach 256 containers for the ``no-op'' function, and 2048 containers for 
the 1-second ``sleep'' function on Theta.
As reported by Wang \emph{et al.}~\cite{wang2018peeking} and Microsoft~\cite{azureFunctionsDocs}, 
for a single function, Amazon Lambda achieves good scalability to more
than 200 containers, Microsoft Azure Functions 
can scale up to 200 containers, and Google Cloud Functions does not scale very well, especially beyond 100 containers.
While these results do not necessarily indicate the maximum number of containers that can be used
for a single function, and likely include some per-user limits imposed by the platform, 
we believe that these results show that \name{} scales similarly to commercial platforms.

\subsubsection{Weak scaling}
To conduct the weak scaling tests we performed \emph{concurrent} function 
requests such that each container receives, on average, 10 requests.
Figure~\ref{fig:scalability}(b) shows the weak scaling for ``no-op,'' 1-second ``sleep,'' and 1-minute ``stress'' functions. 
For ``no-op" functions, the completion time increases with more containers on both Theta and Cori. 
This reflects the time required to distribute requests to all of the containers.
On Cori, \name{} scales to \num{131072} concurrent containers and 
executes more than 1.3 million ``no-op'' functions.
Again, we see that the completion time for 1-second ``sleep'' remains close to constant 
up to 2048 containers, and the completion time for the 1-minute ``stress'' remains 
close to constant up to \num{16384} containers. 
Thus, we expect a function with several
minute duration would scale well to many more containers. 

\subsubsection{Throughput}
We observe a maximum throughput (computed as number of function requests divided 
by completion time) of \num{1694} and \num{1466} requests per second on Theta and Cori, respectively.

\subsubsection{Summary}
Our results show that \name{} 
i) scales to \num{65000}+ containers for a single function; 
ii) exhibits good scaling performance up to approximately \num{2048} containers for a 
1-second function and \num{16384} containers for a 1-minute function; 
and iii) provides similar scalability and throughput using both Singularity and 
Shifter containers on Theta and Cori.

\subsection{Fault Tolerance}
\name{} uses heartbeats to detect and respond to executor failures.
To evaluate fault tolerance we simulate an executor failing and recovering while executing
a workload of sleep functions. To conduct this experiment we deployed \name{} with two executors 
and launched a stream of 100ms functions at a uniform rate such that the system is at capacity.
We trigger a failure of an executor two seconds into the test. 
Figure~\ref{fig:fault_tolerance} illustrates the task latencies measured as the experiment progresses.

We set the heartbeat rate to two seconds in this experiment, causing at least a two second additional 
latency for functions that were inflight during the failure. Following the failure, latencies increase
due to demand exceeding capacity until a replacement executor rejoins the pool, after which task latencies stabilize. 
\kyle{Need to say where this is, and how we get the replacement so quickly}

\begin{figure}[h]
  \includegraphics[width=0.8\columnwidth,trim=0.1in 0.1in 0.1in 0.07in,clip]{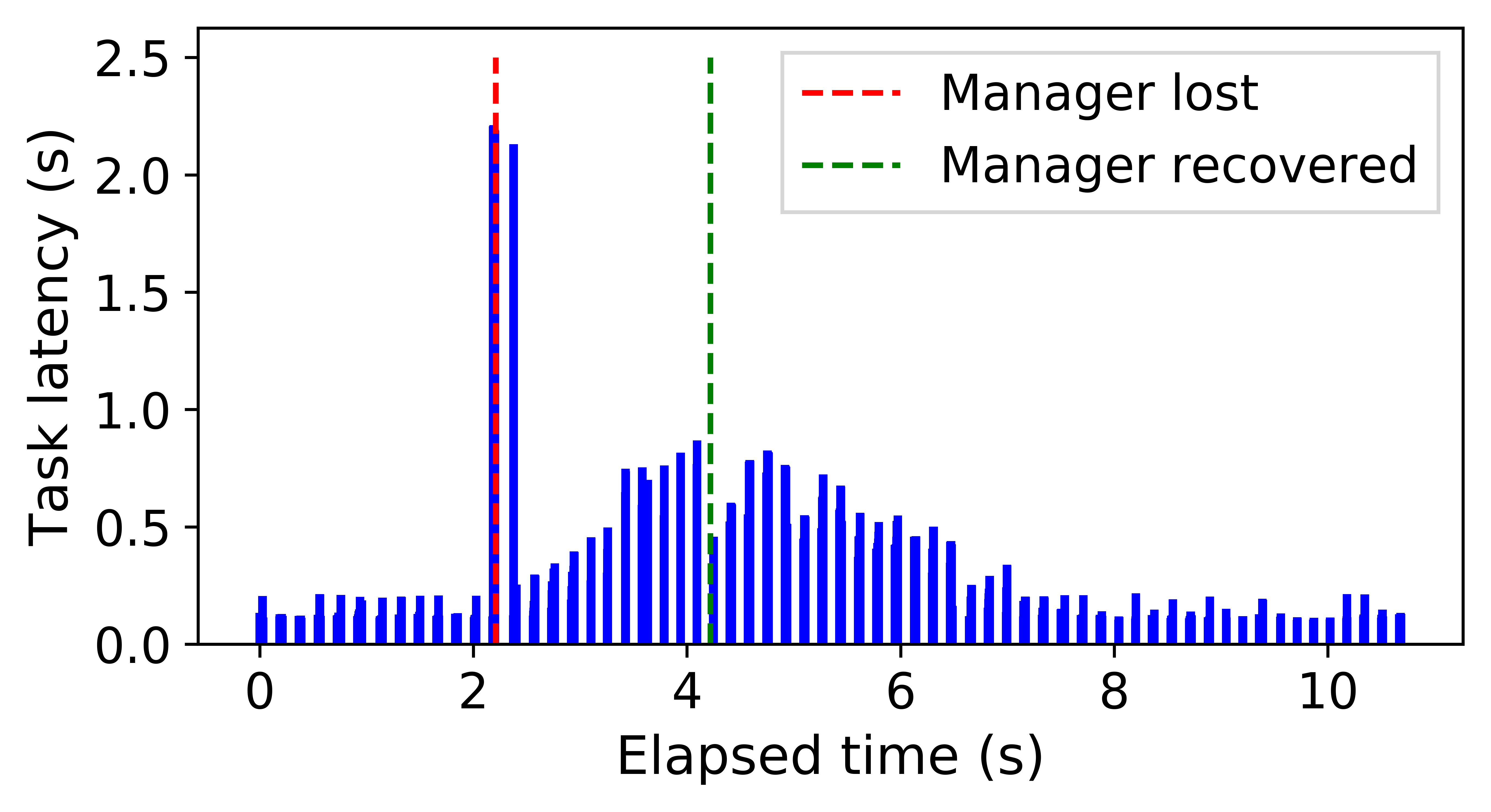}
    \vspace{-0.1in}
  \caption{The latency required to process 100ms functions when an executor fails (2 seconds) and recovers (4 seconds).}
\label{fig:fault_tolerance}
  \vspace{-0.1in}
\end{figure}

\subsection{Optimizations}
In this section we evaluate the effect of our optimization mechanisms. In particular, we investigate how memoization, container initialization, batching, and prefetching impact performance.

\subsubsection{Memoization}
To measure the effect of memoization, we create a function that sleeps for one second 
and returns the input multiplied by two. We submit \num{100000} concurrent function requests to \name{}.
Table~\ref{table:memoization} shows the completion time of the \num{100000} requests 
when the percentage of repeated requests is increased. 
We see that as the percentage of repeated functions increases, the completion time 
decreases dramatically. This highlights the significant performance benefits of memoization
for workloads with repeated deterministic function invocations.
\begin{table}[h]
	\caption{Completion time vs.\ number of repeated requests.}
	\vspace{-0.1in}
	\begin{center}
		\begin{tabular}{c c c c c c}
			\hline
			\textbf{\begin{tabular}[c]{@{}l@{}}Repeated requests (\%)\end{tabular}}
				 & 0 & 25 & 50 & 75 & 100 \\
			\hline
			\textbf{Completion time (s)} & 403.8&  318.5  &    233.6 &  147.9 &   63.2 \\
			\hline
		\end{tabular}
	\end{center}
	\vspace{-0.1in}
	\label{table:memoization}
\end{table}

\subsubsection{Container Instantiation Costs}

To understand the time to instantiate various container technologies on different execution resources
we measure the time it takes to start a container and execute a Python command that imports \name{}'s worker modules---the
baseline steps that would be taken by every cold \name{} function.
We deploy the containers on an EC2 \texttt{m5.large} instance and on compute nodes on Theta and Cori following best practices
laid out in facility documentation. \tablename~\ref{table:cold_start_cost} shows the results.
We speculate that the significant performance deterioration of container instantiation on HPC systems can be attributed
to a combination of slower clock speed on KNL nodes and shared file system contention when fetching images.
These results highlight the need to apply function warming approaches to reduce overheads.

\begin{table}[h]
\caption{Cold container instantiation time for different container technologies on different resources.}
\label{extractor-tab}
 \vspace{-0.1in}
\begin{center}
  \begin{tabular}{l l c c c}
    \textbf{System} & \textbf{Container} & \textbf{Min (s)} & \textbf{Max (s)} & \textbf{Mean (s)} \\
    \hline
    Theta & Singularity & 9.83 & 14.06 & 10.40  \\
    Cori & Shifter & 7.25     & 31.26    & 8.49 \\
    EC2 & Docker   & 1.74     & 1.88     & 1.79 \\
    EC2 & Singularity & 1.19  & 1.26     & 1.22 \\
    \hline
  \end{tabular}
\end{center}
  \vspace{-0.1in}
\label{table:cold_start_cost}
\end{table}

\subsubsection{Executor-side batching}
To evaluate the effect of executor-side batching we submit \num{10000} concurrent ``no-op'' function requests
and measure the completion time when executors can request one function at a time (batching disabled) vs
when they can request many functions at a time based on the number of idle containers (batching enabled). 
We use 4 nodes (64 containers each) on Theta.
We observe that the completion time with batching enabled is 6.7s (compared to 118 seconds when disabled).

\subsubsection{User-driven batching}
We evaluate the effect of user-driven batching we explore the scientific use cases 
discussed in \S\ref{sec:usecases}. These use cases
represent various scientific functions, ranging in execution time from half a second through to almost one minute, and provide perspective to the real-world effects of batching on different types of functions.
The batch size is defined as the number of requests transmitted to the container for execution. 
Figure~\ref{fig:batching} shows the average latency per request (total completion time of the 
batch divided by the batch size), as the batch size is increased. We observe that batching provides 
enormous benefit for the shortest running functions and reduces the average latency dramatically when 
combining tens or hundreds of requests. However, larger batches provide little benefit,
implying it would be better to distribute the requests to additional workers. 
Similarly, long running functions do not benefit as the communication and startup costs are small compared to the computation time.

\begin{figure}[h]
	  \vspace{-0.1in}
	\includegraphics[width=0.8\columnwidth]{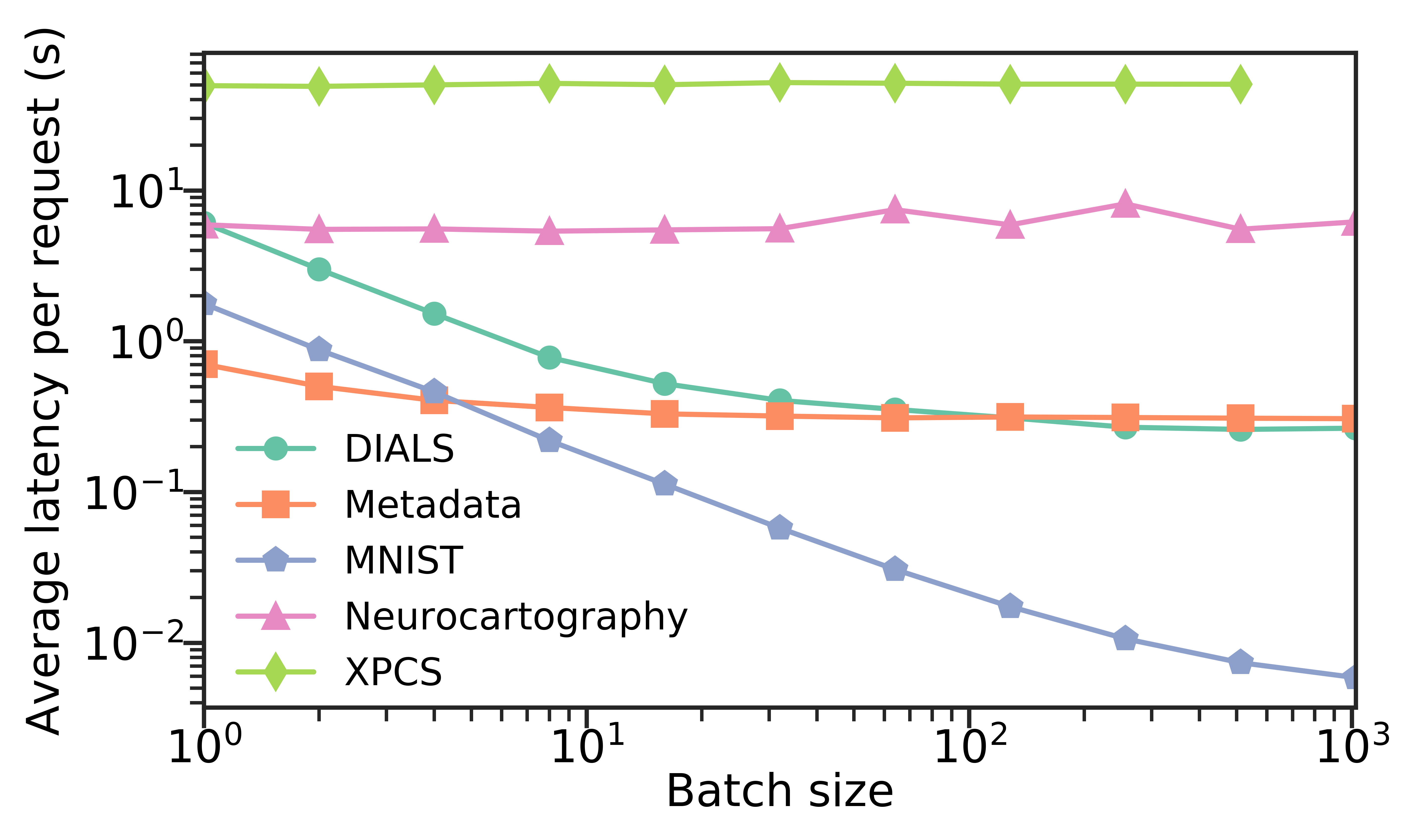}
		  \vspace{-0.1in}
	\caption{Effect of batching on each of the scientific use cases. Batch sizes vary between 1 and 1024.}
	\label{fig:batching}
	  \vspace{-0.1in}
\end{figure}

\subsubsection{Prefetching}
To measure the effect of prefetching, we create ``no-op'' and ``sleep'' functions of different durations (i.e., 1, 10, 100 ms), and measure the completion time of \num{10000} concurrent function requests when the prefetch count per node is increased.
Figure~\ref{fig:prefetching} shows the results of each function with 4 nodes (64 containers each) on Theta. 
We observe that the completion time decreases dramatically as the prefetch count increases. 
This benefit starts diminishing when the prefetch count is greater than 64, which implies that a good setting of prefetch count would be close to the number of containers per node. 
\begin{figure}[h]
	\vspace{-0.1in}
	\includegraphics[width=0.8\columnwidth]{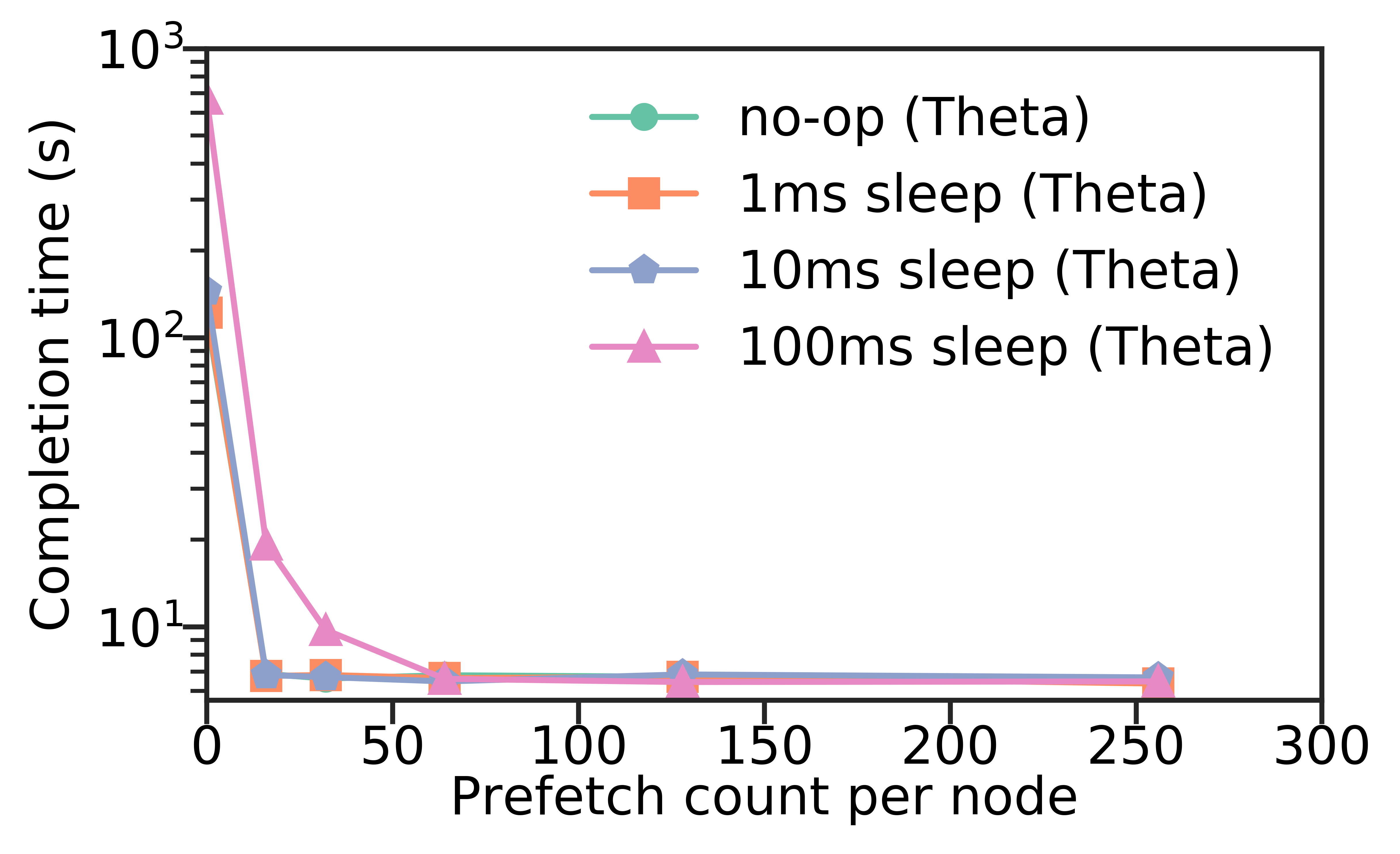}
	\vspace{-0.1in}
	\caption{Effect of prefetching.}
	\label{fig:prefetching}
	\vspace{-0.1in}
\end{figure}

\section{Case Studies}
\label{sec:usecases}

To demonstrate the benefits of \name{} in science we describe
five case studies in which it is being used: scalable
metadata extraction, machine learning inference as a service, synchrotron serial crystallography, neuroscience, and correlation spectroscopy. \figurename~\ref{fig:use-cases} shows execution time distributions for each case study. 
These short duration tasks exemplify opportunities for FaaS in science. 

\begin{figure}[ht]
  \includegraphics[width=1\columnwidth,trim=0.08in 0.1in 0.08in 0.1in,clip]{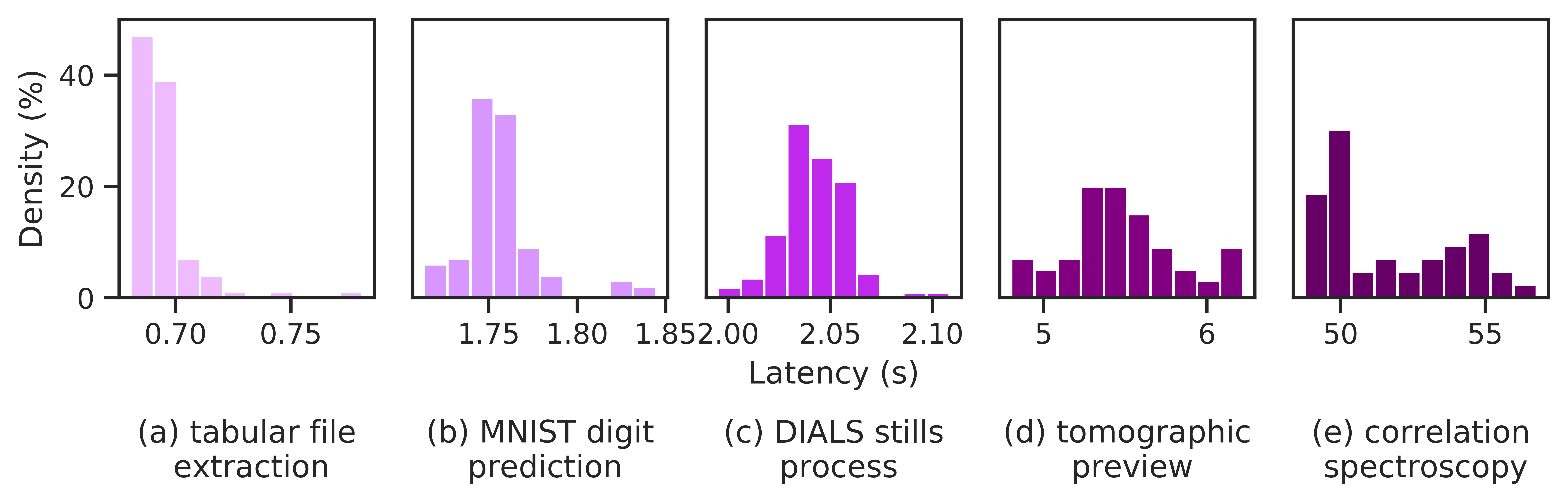}
    \vspace{-0.15in}
  \caption{Distribution of latencies for 100 function calls, for each of the five use cases described in the text.
\label{fig:use-cases}
  \vspace{-0.1in}
}
\end{figure}

\textbf{Metadata Extraction:}
The effects of high-velocity data expansion is making it increasingly
difficult to organize and discover data. Edge file systems and data
repositories now store petabytes of data
and new data is created and data is modified at an alarming rate~\cite{paul17monitoring}.
To make sense of these repositories and file systems, systems such as
\skluma{}~\cite{skluzacek2018skluma} are used to crawl file systems and extract metadata.
\skluma{} is comprised of a set of
general and specialized metadata extractors, such as those designed to process tabular data
through to those that identify locations in maps. All are implemented
in Python, with various dependencies, and each executes for between 3 milliseconds and 15 seconds.
\skluma{} uses \name{} to execute metadata extraction functions directly on the endpoint
on which data reside without moving them to the cloud.

\textbf{Machine Learning Inference:}
As ML becomes increasingly pervasive, new systems are required to support 
model-in-the-loop scientific processes. 
\dlhub{}~\cite{chard19dlhub} is one such tool designed to enable the use of ML in science by supporting
the publication and serving of ML models for on-demand inference.
ML models are often represented as functions, with a set of dependencies
that can be included in a container. 
\dlhub{}'s publication tools help users describe their models using a 
defined metadata schema. 
Once described, model artifacts are published in the \dlhub{} catalog
by uploading the raw model (e.g., PyTorch, tensorflow) and model state (e.g., training data, hyperparameters).
\dlhub{} uses this information to create a container for the model using repo2docker~\cite{repo2docker} that contains all model dependencies, necessary model
state, as well as \name{} software to invoke the model. 
\dlhub{} then uses \name{} to manage the execution of model inference
tasks.  In \figurename~\ref{fig:use-cases} we show the execution time when invoking the 
MNIST digit identification model. While the MNIST model runs for less than two seconds, 
many of the other \dlhub{} models execute for several minutes.
\name{} provides several advantages to \dlhub{}, most notably, that it allows
\dlhub{} to use remote compute resources via a simple interface, and includes performance optimizations (e.g., batching and caching) that improve overall inference performance.

\textbf{Synchrotron Serial Crystallography (SSX)} is a new technique 
that can image small crystal samples 1--2 orders of magnitude faster than other methods~\cite{boutet2012high,wiedorn18megahertz}
and that offers biologists many new capabilities, such as imaging of conformation changes,
very low X-ray doses for sensitive samples, room temperature for more biologically relevant environments, 
radiation sensitivity for metalloproteins, and discovery of redox potentials in active sites. 
To keep pace with the increased data production, SSX researchers require new automated methods
of computing that can process the resulting data with great rapidity:
for example, to count the \emph{bright spots} in an image 
(``stills processing'') within seconds, both for quality control 
and as a first step in structure determination.
We have deployed the DIALS~\cite{waterman2013dials} crystallography processing tools as 
\name{} functions. \name{} allows SSX researchers to submit the same \emph{stills process} function to either a local endpoint to perform data validation or offload large batches of invocations to
HPC resources to process entire datasets and derive crystal structures.

\textbf{Quantitative neurocartography} and connectomics involve the mapping of the 
neurological connections in the brain---a compute- and data-intensive processes 
that requires processing \textasciitilde 20GB every minute during experiments.
We have used \name{} as part of an automated workflow
to perform quality control on raw images (to validate that the instrument
and sample are correctly configured), apply ML models
to detect image centers for subsequent reconstruction, and generate
preview images to guide positioning. 
\name{} has proven to be a significant improvement over previous practice, which
depended on batch computing jobs that were subject to long scheduling delays
and required frequent manual intervention for authentication, configuration, and failure resolution. 
\name{} allows these workloads to be more flexibly implemented, making
use of a variety of available computing resources, and removing overheads of
managing compute environments manually. 
Further, it allows these researchers
to integrate computing into their automated visualization and analysis workflows 
(e.g., TomoPy~\cite{gursoy2014tomopy} and Automo~\cite{Automo}) via programmatic APIs.

\textbf{X-ray Photon Correlation Spectroscopy (XPCS)} 
is an experimental technique used at Argonne's Advanced Photon Source
to study the dynamics in materials at nanoscale by identifying correlations in 
time series of area detector images. 
This process involves analyzing the pixel-by-pixel correlations 
for different time intervals. 
The current detector can acquire megapixel frames at
60 Hz (\textasciitilde 120 MB/sec). 
Computing correlations at these data rates is a challenge that requires HPC resources but also
rapid response time. We deployed XPCS-eigen's corr function as a \name{} 
function to evaluate the rate at which data can be processed. Corr is able to process a dataset 
in \textasciitilde 50 seconds. Images can be processed in parallel using \name{} to invoke corr functions on-demand.

\textbf{Lessons learned:}
We briefly conclude by describing our experiences applying \name{} to the five
scientific case studies. Before using \name{}, these types of use cases
would rely on manual development and deployment of software on 
batch submission systems.

Based on discussion with these researchers we have identified the following 
benefits of the \name{} approach in these scenarios. 
First, \name{} abstracts the complexity of using HPC resources. 
Researchers were able to incorporate scalable analyses using
without having to know anything about the computing environment 
(submission queues, container technology, etc.) that was being used. 
Further, they did not have to use cumbersome 2-factor authentication, manually scale workloads, or map their applications to batch jobs. 
This was particularly
beneficial to the SSX use case as it was trivial to scale the analysis
from one to thousands of images. 
Many of these use cases use \name{} to enable event-based processing. 
We found that the \name{} model lends itself well to such use cases, 
as it allows for the execution of sporadic workloads. For example, 
the neurocartography, XPCS, and SSX use cases all exhibit such characteristics, 
requiring compute resources only when experiments are running. 
Finally, \name{} allowed users to securely share their codes, allowing 
other researchers to easily (without needing to setup environments) apply
functions on their own datasets. This was
particularly useful in the XPCS use case as many researchers share
access to the same instrument.

While initial feedback has been encouraging, our experiences also highlighted 
several challenges that need to be addressed. 
For example, while it is relatively easy to debug a running \name{} function, it 
can be difficult to determine why a function fails when first published. 
Similarly, containerization does not necessarily provide entirely portable 
codes that can be run on arbitrary resources due to the need to compile and link resource-specific modules. For example, in the XPCS use case
we needed to compile codes specifically for a target resource. 
Finally, the current \name{} endpoint software does not yet 
support multiple allocations. To accommodate more general use
of \name{} for distinct projects we need to develop a
model to specify an allocation and provide accounting and
billing models to report usage on a per-user and per-function basis.

\section{Conclusion}\label{sec:conclusion}

Here we presented \name{}---a FaaS platform designed to enable the low latency, scalable, and secure execution
of functions on almost any accessible computing resource. \name{} can be deployed on existing HPC infrastructure
to enable ``serverless supercomputing.'' We demonstrated that \name{} provides comparable latency 
to that of cloud-hosted FaaS platforms and showed that \name{} can execute 1M tasks
over \num{65000} concurrent workers 
when deployed on 1024 nodes on the Theta supercomputer. 
Based on early experiences using \name{} in five scientific use cases we have found that the approach
is not only performant but also flexible in terms of the diverse requirements it can address. 
In future work we will extend \name{}'s container management capabilities to dynamically 
create containers based on function requirements and stage them to endpoints on-demand.
We will also explore techniques to share containers between functions with similar dependencies. 
We plan to design customized, resource-aware scheduling algorithms to further improve performance. 
Finally, we are actively developing a multi-tenant endpoint and the additional isolation techniques 
necessary to provide safe and secure execution. 
\name{} is open source and available on GitHub.

\section*{Acknowledgment}
This work was supported in part by Laboratory Directed Research and
Development funding from Argonne National Laboratory under U.S. Department of
Energy under Contract DE-AC02-06CH11357. This research used resources of the
Argonne Leadership Computing Facility, which is a DOE Office of Science User
Facility supported under Contract DE-AC02-06CH11357. We thank the Argonne
Leadership Computing Facility for access to the PetrelKube Kubernetes cluster
and Amazon Web Services for providing research credits to enable rapid service
prototyping.

\bibliographystyle{ACM-Reference-Format}
\bibliography{Bibs/refs,Bibs/funcx-refs,Bibs/newrefs,Bibs/refs-old,Bibs/ipdps19}

\end{document}